\documentclass[10pt,letterpaper]{article}
\usepackage{epsfig}

\begin{document}

\title{\large \bf Semantic Network Layering}
\author{Michael Neufeld and Craig Partridge \\
BBN Technologies \\
10 Moulton St. \\
Cambridge, MA 02138 \\
mneufeld@bbn.com, craig@bbn.com}

\maketitle

\begin{abstract}
The stack in various forms\cite{osi:zimmerman} has been widely used as
an architectural template for networking systems. Recently the stack
has been subject to criticism for a lack of flexibility.
However, when it comes right down to it nobody
has offered a truly compelling alternative. Various ``cross-layer''
optimizations have been proposed, but these optimizations are frequently
hacks to achieve a particular goal and offer no direct insight into
{\em why} the existing network stack is inadequate. 
We propose that a fundamental problem with the existing network
stack is that it attempts to layer functionality that is not 
well-suited to layering. In this work we use a ``bottom up''
model of information computation, storage, and transfer 
and the ``top down'' goals of networking systems to formulate
a modular decomposition of networking systems. Based on this
modular decomposition we propose a semantic layered structure for
networking systems that eliminates many awkward
cross-layer interactions that arise in the canonical layered
stack.
\end{abstract}

\section{Introduction}
\label{sec:intro}
As a starting point we will define what exactly networking
and storage systems do. Stated simply,
the goal of networking and storage is to move symbolic
information to specified places and times without altering the meaning
of that symbolic information, subject to some set of
constraints on resources and operational parameters.
In any real system moving an abstract symbol entails 
moving the {\em embodiment} of
that symbol. When a symbol is not embodied in a form suited
for travel the symbol must first be transferred to 
some embodiment that {\em is} suited for travel, moved, and
then transferred into whatever embodiment is required at the
destination. We will explore what makes for a ``suitable'' embodiment
for information in Section~\ref{section:netstorecomp}.
Constraints may apply to
concrete physical {\em resources}, {\em e.g.} energy
and raw materials, subjective and abstract resources, {\em e.g.}
social standing and opportunities when interacting with other
entities, as well as on {\em actions}, {\em e.g.} a wireless system
may only be allowed to operate on particular frequencies at particular times.
We will explore goals, actions, resources, and constraints in greater detail in
Section~\ref{section:actionsresourcesgoals}.
In Section~\ref{section:networkfunctions} we will further
break down the task of moving information in
space and time into multiple sub-tasks, defining the functions
that networking and storage systems must perform.

\section{Networking, Storage, and Computation}
\label{section:netstorecomp}
In this section we will broadly categorize networking,
storage, and computation. This categorization is based
on fundamental tenets of modern physics, and was constructed under the
influence of work by Feynman\cite{feynman}, Zuse\cite{zuse} and
Fredkin\cite{fredkin} regarding fundamental notions of computation
and physics. We do not claim that this is a novel categorization, and this
work is not primarily about computation and storage. However, we feel that
networking, storage, and computation are all fundamentally related
tasks, and that having a firm notion of all three is
critical for contextualizing and understanding networking.
In the universe as we understand it computationally useful symbolic
information must reside within some physical embodiment. This embodiment
may be a wide range of things, {\em e.g.} DNA and RNA within cells,
electrical potential, or a quantum superposition of spin states.
\begin{figure}
\centerline{
\begin{tabular}{ccc}
\epsfig{file=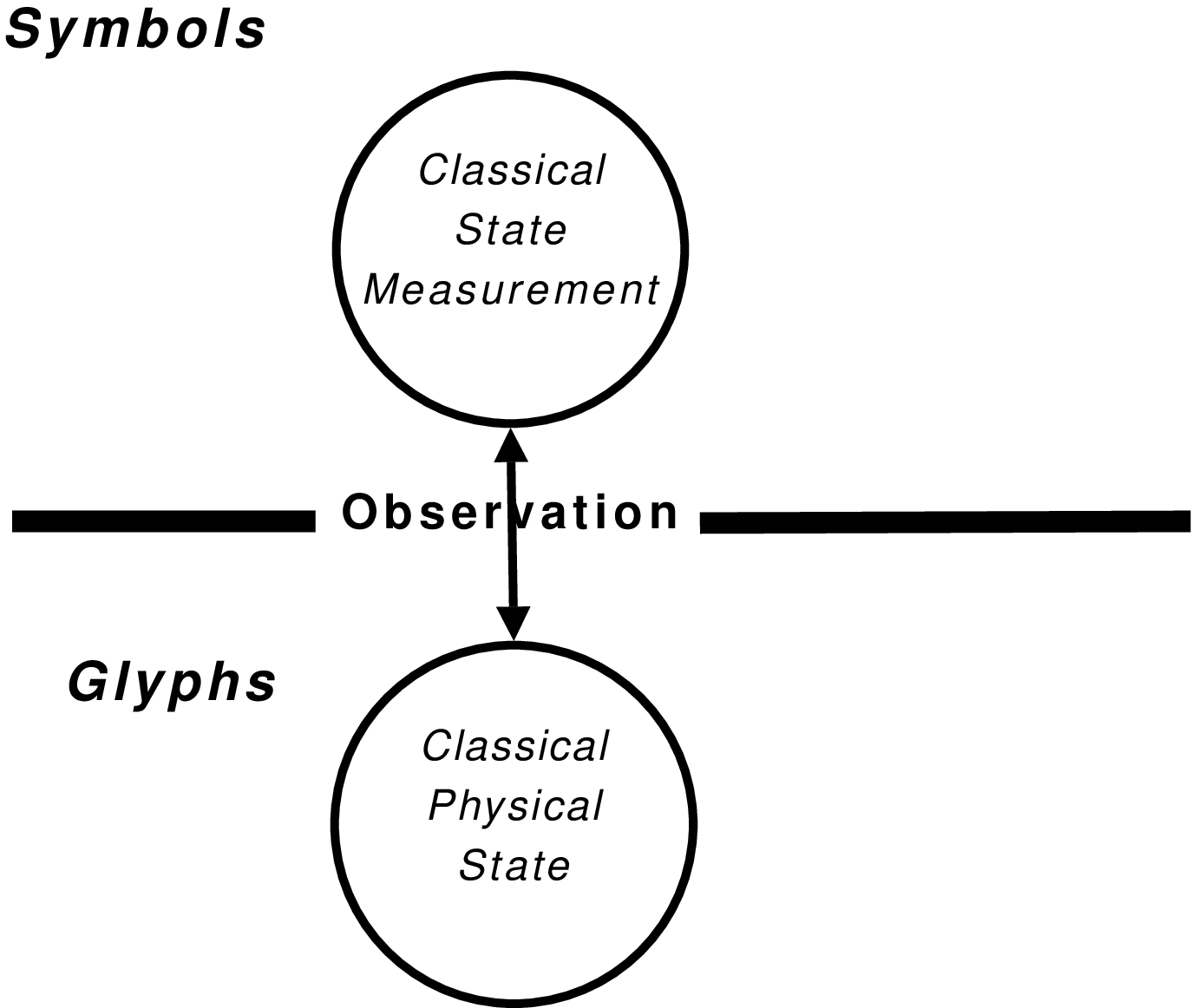,width=0.3\columnwidth}
&
\epsfig{file=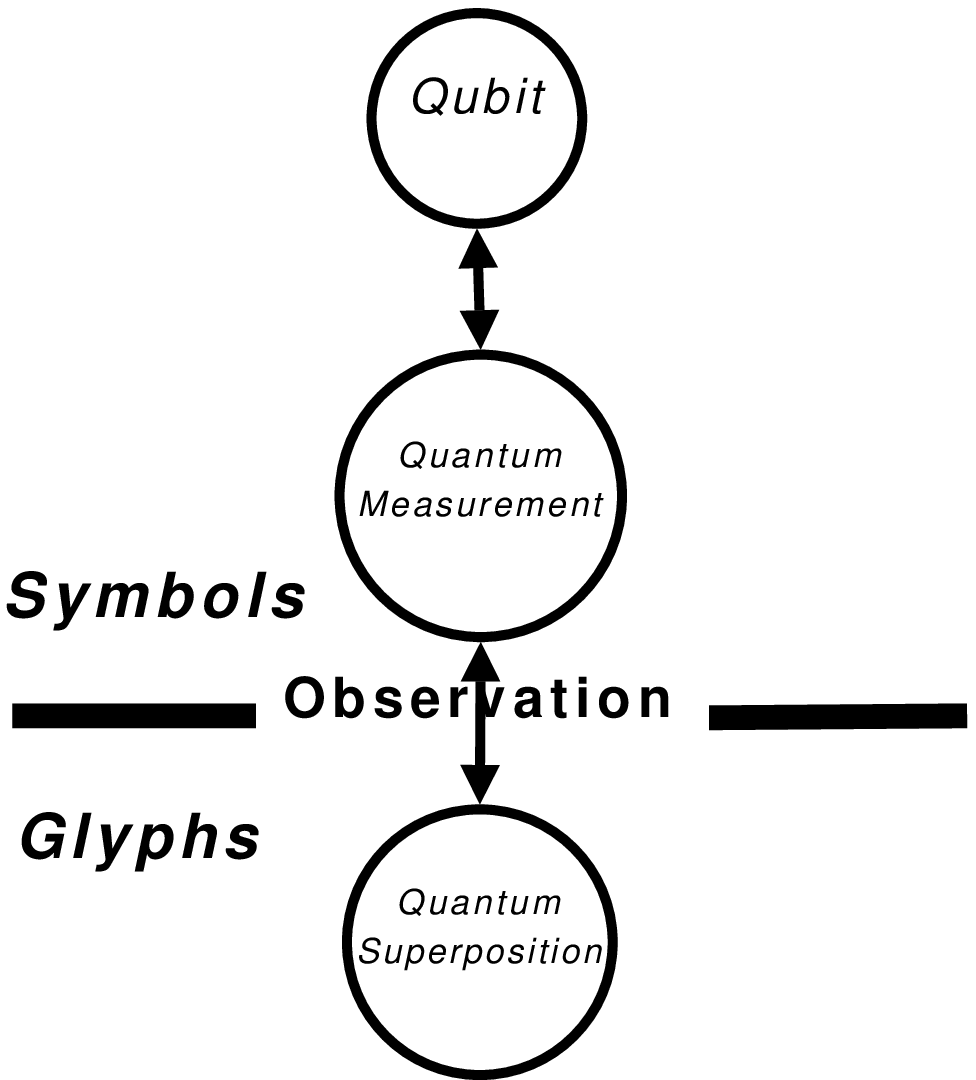,width=0.3\columnwidth}
&
\epsfig{file=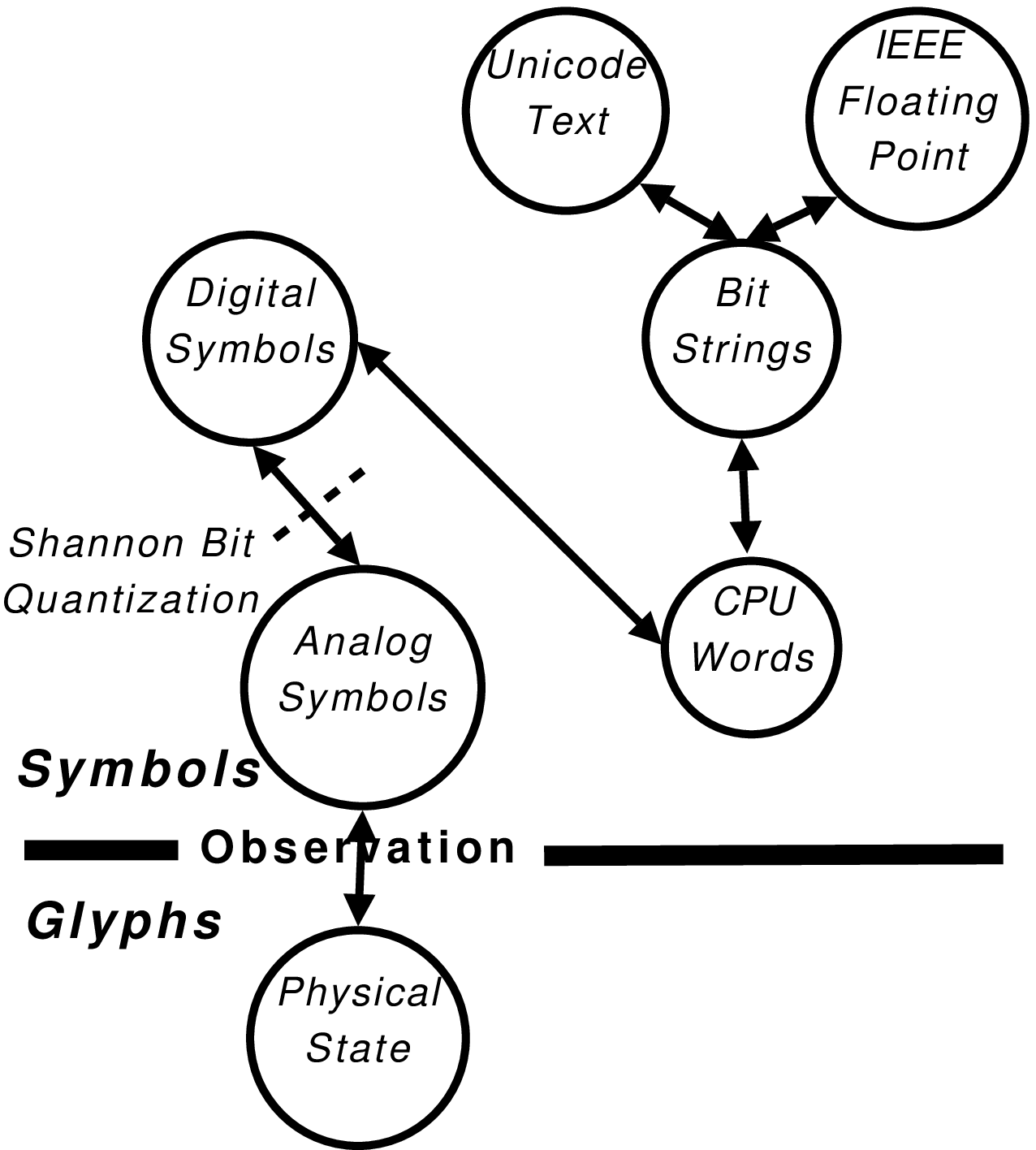,width=0.3\columnwidth}
\\
\begin{minipage}[c]{0.3\columnwidth}
\centerline{
\footnotesize Analog
}
\end{minipage}
&
\begin{minipage}[c]{0.3\columnwidth}
\centerline{
\footnotesize Quantum
}
\end{minipage}
&
\begin{minipage}[c]{0.3\columnwidth}
\centerline{
\footnotesize Digital
}
\end{minipage}
\end{tabular}
}
\caption{\footnotesize A diagram showing how embodiments, symbols, and filtering combine to create symbol sets for a analog, quantum, and traditional digital computation systems.}
\label{figure:symbol_systems}
\end{figure}
Figure~\ref{figure:symbol_systems} shows diagrams of how
symbols for analog, quantum, and digital computation system are formed from
embodiments. In the digital system, glyphs are measured to create analog
symbols, which are then further processed to generate digital symbols, and
then mapped into bit symbol
strings of machine word size. These machine words may be
further mapped into operators of a microprocessor instruction set, Unicode
characters of human language, or any number of other symbolic domains.
The interpretation and semantics of symbols has been formally
studied as {\em denotational semantics}\cite{denotationalsemantics},
and semantics are central to computation, networking,
and storage systems.

\begin{figure}
\centerline{
\begin{tabular}{cc}
\epsfig{file=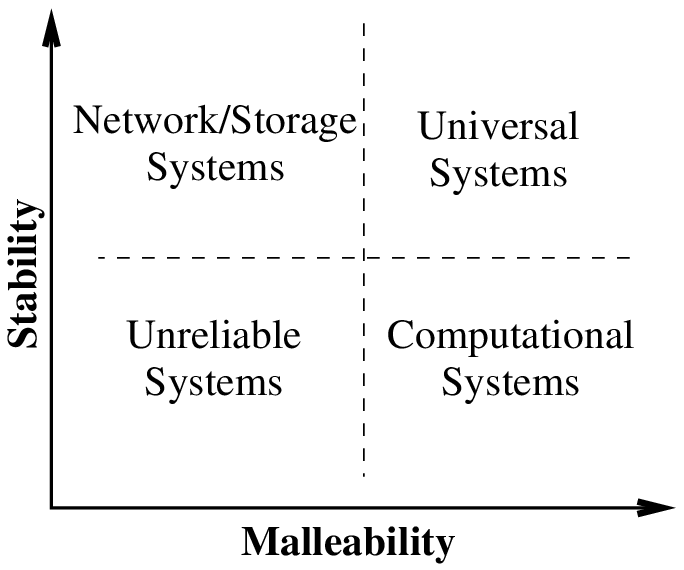,width=0.4\columnwidth}
&
\epsfig{file=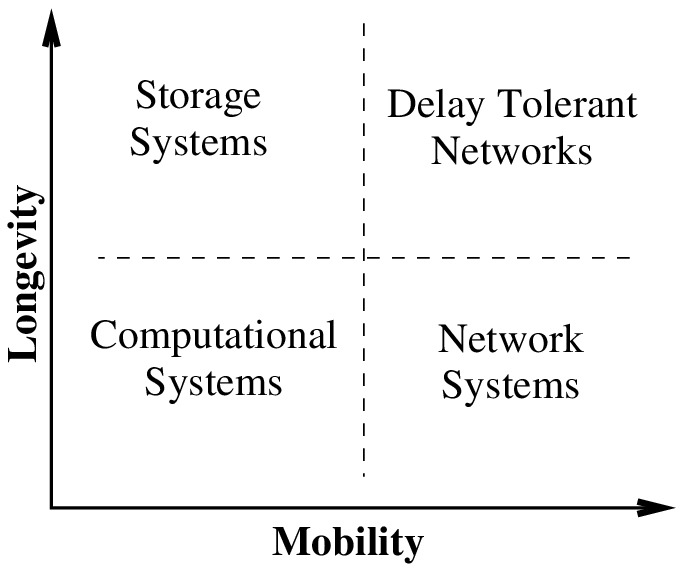,width=0.4\columnwidth}
\\
\begin{minipage}[c]{0.45\columnwidth}
\centerline{
\footnotesize Stability and Malleability
}
\end{minipage}
&
\begin{minipage}[c]{0.45\columnwidth}
\centerline{
\footnotesize Mobility and Longevity
}
\end{minipage}
\end{tabular}
}
\caption{\footnotesize Computational, networking and storage systems place different demands on glyphs. Stability and malleability measure the difficulty and speed of glyph alteration respectively. Mobility and longevity measure how difficult it is to move the glyph in space and time respectively.}
\label{figure:compnetstore}
\end{figure}

Of course, not all embodiments are alike, and some embodiments
may be more suited to some purposes than others. We will
broadly categorize ``purposes'' into three categories:
networking, computation, and storage. Computation generally requires
a low amount of travel in time and space, but an ability to rapidly
change state when desired. Networking and storage must both
be stable and able to survive over whatever lengths of time and
space are required.
Choosing an embodiment for networking, storage, or computation
is fundamentally motivated by physical system characteristics.
From a general system architecture design standpoint
we propose four critical physical properties: {\em stability}, {\em i.e.}
the difficulty/energy required to change a state, {\em malleability},
{\em i.e.} how quickly it may be changed, {\em longevity}, {\em i.e.}
how easily it may be moved through time, and {\em mobility}, 
{\em i.e.} how easily it may be moved through space. These properties are
illustrated in Figure~\ref{figure:compnetstore}. 
Expressing these
properties quantitatively relies on an ability to handle physical space,
time, and energy. Because networking and storage systems strive
to preserve symbolic state we are also interested in {\em distortion},
{\em i.e.} unintended changes in physical and/or symbolic state,
and {\em fragmentation} when symbols do not fit into a single
embodiment. When treating an embodiment as a measured time-varying
signal the {\em bit}\cite{shannon-bit} may be used to quantify and
analyze distortion as well as the natural size of an underlying
symbol embodiment. The bit may also be used in conjunction with
coding theory to design {\em translation} schemes that may be
used to stave off symbol distortion.
Unfortunately the term ``bit'' is overloaded: a ``bit'' not
only refers to a specific way of converting a noisy signal into
discrete symbols, but also the use of binary digits as a universal
symbolic alphabet. When working with physically discretized systems
it is convenient to use symbolic bits to represent states, 
but this is separate from performing bit-style signal filtering and
discretizing.

\section{Goals, Actions, Resources, and Constraints}
\label{section:actionsresourcesgoals}
Any practical system will have specific {\em actions} that
it may take as part of its operation, {\em resources} that are
required to take action, {\em constraints} placed on actions
and resources, and some fundamental set of {\em goals} that
it should accomplish. In this section we will describe
goals, resources, and constraints of networking systems
in the context of both physical and symbolic computation
(we will discuss {\em actions} later as part of
our functional modularization of networking systems).
In addition, we will also examine goals, resources,
and constraints for communication systems in the {\em human} domain.
The human domain is concerned with how humans build, operate,
and use computational systems. Such a perspective is vital
for building systems that are to be adopted by human society at large,
and we will use this perspective as one of two top down
perspectives.  Our
second top down perspective is a generic symbolic view of computation
and communication. Our bottom up perspective
is a physical view of computation and communication. The human and
symbolic domains are connected through the physical world
via {\em human computer interaction} (HCI) devices and applications,
and different symbolic computation domains are connected through the
physical world via networking devices and applications. In this work we
will ignore specific details of HCI devices and applications, but
still attempt to take the general demands of human users into account
when determining what features a symbolic networking system should
have.

We will broadly categorize goals, resources,
and constraints into two major areas: {\em physical}
and {\em social}. Physical elements include computation, 
but are more fundamentally concerned with the physical world,
{\em e.g.} energy, space, time, and raw materials. 
Social elements are those which arise from interactions
between multiple autonomous entities, and not obviously derived
from physical or chemical first principles. Artificial
social interactions may often be analyzed mathematically,
{\em e.g.} carrier sense multiple-access systems\cite{kleinrock1}.
Biological social systems are not generally amenable to
precise quantification, though economists have made extensive
efforts to create mathematical models capturing aspects
of human commerce with varying degrees of success.
Social and physical goals are often 
intertwined and interdependent. For example, people may cooperate socially
in order to hunt, gather, and farm food (energy) more effectively,
and conversely individuals may strive to hunt, gather, and farm more
effectively in order to increase social standing. We will not
engage in an extended discussion about the
precise distinction between social and physical aspects of
systems, but we will emphasize that social aspects will arise
and have to be addressed in networking systems.

\subsection{Goals}
\label{section:goals}
The physical goals of a networking system within the human domain
generally concentrate on delivering media to
some set of people (or other creatures capable of communication)
with some specified acceptable amount of distortion and error
within some specified amount of time. These physical goals
are often driven by underlying social goals. For example,
the social goals of security and privacy may dictate that some set of
entities be unable to intercept and/or interpret the information
during transit, potentially placing physical constraints on operation.
For symbolic computation, the physical goal of a networking
system is to deliver specified sets of symbols to locations where
further computation, storage, or transfer of those symbols may occur.
Physical computation is tasked with moving some collection
of physical embodiments to some specified locations in space and time
with some limit on the allowed distortion of those embodiments.

Social goals for symbolic and physical computation are
built in to a system by its designers, and often reflect
the social goals of the designers, users, and builders in the
human domain. {\em Fairness} is used as a metric for evaluating
network protocols, and {\em Dynamic Spectrum Access} (DSA)\cite{dsaoverview}
techniques usually strive to be ``good neighbors'' by only utilizing
unoccupied radio frequencies and obeying usage laws laid out by
governmental agencies. The current debate over ``network neutrality''
and traffic shaping further illustrates how human social factors
figure in to protocol design. 

\subsection{Actions}
\label{section:actions}
The specific actions that a system make take vary depending
on the system. In this section we will identify the
broad actions that communication systems may take. 
Two key actions taken on information are to {\em transmit} information over
some region of space and time, and to {\em receive} information
from some region of space and time. Coordinating the transmit and
receive operations such that they overlap in space, time, and
physical embodiment is central to communication path establishment,
or {\em linking}. Linking does not always operate directly on information,
{\em e.g.} autonomous robotic systems may {\em move} in order to enhance
communication\cite{netmotion} or alter their configuration to facilitate
communication, {\em e.g.} reconfigurable directional antennas.
Advanced automated systems could even construct their own
physical infrastructure for communication, {\em e.g.} deploying 
wire or fiber for fixed infrastructure, or even microscopic filaments
for nanobots\cite{molecular}.
{\em Translation} of information into another form,
{\em e.g.} coding for distortion resistance and spatial/temporal footprint
and {\em buffering} packets for flow control are information-centric
actions that may be used for linking. Further information computation and
storage operations are {\em copying} and {\em erasing}.
Copying information may be very time and energy intensive in
conventional systems, and expressly forbidden for quantum information.
For reversible systems the act of {\em erasing} information is of
fundamental significance, and may be expensive in conventional
systems. We will discuss further details of these actions when
we engage in a more detailed exploration of networking system
architectural details.

\subsection{Physical Resources and Constraints}
\label{section:physicalresources}
Physical resources and constraints are, essentially, about
time, space, energy, and entropy. As might be expected, physical
computation is directly aware of physical constraints on
on space, time, and energy. The human domain must also account
for physical space, time, and energy. Humans, in general, care
about where and when computation occurs, and may have to manage
limited energy resources.
Symbolic computation may subsume some
physical constraints into more abstract constraints on network, storage, and
computational capacity, though physically situated symbolic applications,
such as sensor networks and autonomous robotic systems, may have extensive
and detailed models of the physical environment. Furthermore, the symbolic
computation domain should have some notion of space, time, and energy
if it is to interact naturally with the human domain. Most conventional
networking systems have striven to abstract away physical time and space
through logical clocks and virtual coordinate systems. Such abstraction
is useful in that it reduces reliance on clock synchronization and
location knowledge. However, the complete removal
of physical time, space, and energy constraints can make it awkward
to connect the symbolic computation domain with both the physical
computation domain and the human domain. 

\subsection{Social Resources and Constraints}
\label{section:socialresources}
{\em Social} constraints exist when multiple entities interact.
In the realm of physical computation, MAC layers use various
social strategies ({\em protocols}) to control access to shared
communication media, such as {\em carrier sense} (transmit only after
you can't hear someone else transmitting), and dividing up a channel
into frequency or time slices (frequency/time division multiple access).
Traffic shaping, firewalls, and inter-AS forwarding are all examples
of social strategies for symbolic computation systems.
Social strategies for automated physical and symbolic systems are
generally set within the context of some mathematical analysis of a
system model, {\em e.g.} CSMA for wireless systems\cite{kleinrock1,kleinrock2}.
The influence of the human domain is imprinted on the social structures
of physical and symbolic computation systems. 
Protocol design usually involves extensive committee work\cite{80211spec},
issues of ``fairness'' arise when designing new or
modifying previously-deployed access schemes for networks,
and game theory is periodically applied in attempts to design socially
sustainable automated interacting systems.
Furthermore, human laws may directly influence physical and symbolic
systems directly, {\em e.g.} dynamic spectrum access\cite{dsaoverview}
systems that comply with government regulations on RF spectrum usage.
Social resources and constraints within the human domain manifest
directly within applications. The use of e-mail, Usenet, chat, peer-to-peer
file distribution services, and social networking services all have
developed social protocols. {\em Spam} is a primary example of what happens
when voluntary social constraints are ignored, resulting in surrounding
neighbors instituting other social constraints (blacklisting, filtering,
{\em etc.}) as mitigating measures.

\section{Network Functionality}
\label{section:networkfunctions}
Given the goals, constraints, and resources that we have enumerated
for networking systems, how can we best partition functionality
out into modules? In Section~\ref{section:netstorecomp} we identified
three functions that arise from considering the physical
embodiment used for symbols: fragmentation, distortion control,
and symbol translation. Another core task is {\em identification},
{\em i.e.} naming and addressing. Networking systems need to have
some way of identifying and locating symbols and endpoints if they
are to manage the motion of symbols and endpoints. We will also
break down the motion of symbols and endpoints into separate tasks:
{\em topology control} for managing endpoints and symbol paths between
endpoints and {\em flow control} for the managing the motion of symbols.
The final function we will identify is {\em multiplexing}, {\em i.e.}
allowing multiple clients to share a single underlying service,
which has been empirically identified as useful for networking
systems\cite{alf}. Naturally, this is not the only possible decomposition
of networking into functions, though it is generally similar to
other decompositions\cite{alf,arch:rfc} in many important respects.
All of these functions may interact with each other, potentially
in arbitrarily complex ways. A good architecture can help
mitigate this complexity by making some core decisions about
interactions in a way that balances abstraction, transparency,
flexibility, and efficiency in a useful way. Of course, this
task is much easier said than done, and has been a topic of practical and
academic study for many years. 
In this section we will explore each of the seven functions we have
identified as well as how those functions interact with each other
and the constraints placed upon them.

%
%
%
%
%
%
%

\subsection{Identification: Naming and Addressing}
\label{section:identification}
Earlier work\cite{names-hauzeur} has stressed the fundamental nature
of names, addresses, and routes. Before we define names and addresses
we will first examine the broader problem of {\em identification}.
Deep philosophical discussions aside, we propose that identification
can be broadly split into two major components: {\em location} and
{\em appearance}. Location is simple location in time and space, appearance
is some set of observed properties of an entity. Appearance may involve
aspects of location, but we will require that location-based elements of
appearance be expressed within some local coordinate system of the entity
being identified. For example, 3D polygonal models of objects are
usually constructed in reference to some local object center and
axes that may then be translated, rotated, and distorted as required
to embed the model in larger scenes.
Conversely, coordinate systems used for specifying location may rely on
appearance to define critical points, {\em e.g.} the North and South Magnetic
Poles of the Earth. Fortunately, human civilization has devised many
useful coordinate systems, relieving most systems from the
burden of devising their own. With this in mind, we will not
discuss the fundamental philosophical issues associated
with defining coordinate systems, instead continuing on with practical
aspects of identification. We will generically refer to an observed
set of properties of an entity as a {\em label}. When a label
specifies appearance we will call it a {\em name} and when a label
specifies location we will call it an {\em address}. Identification
is then achieved through some set of labels and/or constraints
on labels.

Physical computation uses correspondingly
physical names and addresses. For example, directional antenna
systems use physical directions for addressing and
photon emission patterns for naming.
The human domain utilizes both physical and symbolic
schemes for naming and addressing, {\em e.g.} GPS coordinates
and physical descriptions of size and color as well as names
for abstract concepts, geographic regions, people, and other entities.
Symbolic computation names and addresses can be very abstract
as well. Existing network systems often take a very
abstract ``mathematical'' view of naming and addressing.
For example, DNS, IP, and Ethernet identifiers are static strings of symbols.
IP addresses reflect network topology, and may be correlated only
loosely and incidentally with geographic location. IP addresses
are also widely used as names, causing quite a bit of extra
complication when a network attachment point changes\cite{mobileip:rfc}.
Such abstract symbolic systems may run into difficulty when attempting
to mediate between the physical domain and the physically-oriented
parts of the human domain. Attempting to eliminate
physical considerations from the symbolic domain entirely can
result in awkwardness when the human domain wants to utilize
time and location information from the physical domain and vice versa.

An important question for any system is what kinds of entities require
identification. For networking systems we will identify three major
entity types: {\em endpoints}, {\em information}, and {\em paths}.
The precise definitions of endpoint, information
and path depend on the particular system in use and the level
of abstraction at which the system operates. In the human domain, 
people are quite often the intended endpoints, generally interacting
with some computational application as a proxy.
Information may be any content used by humans (text, video,
audio, still pictures, {\em etc.}), and paths typically consist
of the people and places through which and to which the information flows.
In the traditional ``operating system'' view of symbolic
computation, applications are the endpoints, information resides
within files, and paths are ``symbol pipes'' between applications
with some (possibly time-varying) capacity and latency.
For physical computation the endpoints may be any physical entities
that interact with the information embodiment, information consists
of some collection of physical embodiments, and paths are defined
in terms of physical space and time as well as the physical
entities (endpoints) that interact with the embodiments as they travel.

\subsection{Topology Control}
\label{section:topologycontrol}
{\em Topology Control} tracks and manages endpoints and the paths between
endpoints where information may travel. 
For physical computation topology management is tightly tied to the
physical world:
endpoints and paths are physically identified regions in time and space,
though they may also have names associated with them that directly
relate to physical properties. Controlling topology in this context
means directly controlling aspects of the physical environment as
well as coordinating control of the physical environment with other
entities. 
Topology control in the human domain often explicitly involves physical
time and space, but also has a significant component related
to named entities. Paths are likely to be defined in terms
of {\em who} can receive the information at what times as well
as physically {\em where} the information goes at what times.
Symbolic computation has the option of ignoring
space and time completely, {\em e.g.} viewing endpoints and
paths as abstract graphs. However,
as we briefly discussed in Section~\ref{section:identification}
it is generally sensible for the symbolic domain to maintain
some models of physical time and location in order to mediate
more effectively between the physical and human domains.

\subsection{Flow Control}
\label{section:flowcontrol}
{\em Flow Control} moves and tracks information through time and space
via the paths determined by Topology Control.
As such, Flow Control is the ``prime mover'' of a networking
system. Without Flow Control, Topology Control would not
know where best its paths should go. Distortion Control and
Fragmentation/Reassembly would have no moving information
on which to operate. The primacy of Flow Control is obscured
in many systems by the complexity of Topology Control. 
Many traditional routing systems attempt to preemptively discover and track
{\em all possible} destinations and paths in a network, allowing
Flow Control to select any destination within the network as desired.
Furthermore, Flow Control is not usually given any choice of routes
taken by an individual packet. Multihoming may
be exploited to encourage multipath routing\cite{sctp-multipath},
but such schemes could clearly benefit from having a more direct
and overt way to exploit multipath diversity. An important feature of
our semantic stack is that it does not mandate {\em layering} as the primary
modularization of Flow Control and Topology Control, as in a 
stack based on functional layers.

\subsection{Distortion Control}
\label{section:distortioncontrol}
{\em Distortion Control} ensures the integrity of information
by controlling unwanted changes to information. For physical computation
distortion is tied directly to physical attributes of embodiments,
{\em e.g.} the frequency and polarization of a photon. As such
it is important to characterize the physical environment 
({\em i.e.} the channel) and its effect on the physical embodiment.
For example, radio waves in a highly reflective environment
often experience multipath distortion at the receiver. One way
of mitigating multipath distortion at the physical level would be
to embed symbols in the frequency of the photons and not the phase
so that the symbols directly measured from the information
embodiment experienced less distortion in the channel. Another
option would be to encode the information in such a way that
the low-level physical symbol distortions did not result in higher-level
information symbol distortions, or even {\em exploit} the multipath
reflections (as with {\em MIMO} systems).
Such {\em channel coding} techniques
are a fundamental part of information theory.
Symbolic computation has a more abstract notion of distortion than
physical computation. 
Physical distortion and its immediate effect on symbolic interpretation
is not modelled in detail. Instead, distortion occurs as alterations to
symbols or the loss of symbols entirely, {\em e.g.} bit errors and loss
in binary symmetric channels and binary erasure channels.

Moving into the human domain the notion of distortion becomes
even more abstract. For example, a sequence of numbers may be interpreted
as a sequence of characters used in human languages, such as EBCDIC, ASCII,
and Unicode. These strings may be further interpreted as words in human
languages. In isolation it may not be clear which language
is intended. For example, the word ``die'' on its own could
be a word in German or English, meaning entirely different
things to speakers of those languages. Even within a single language
words and sentences may be semantically ambiguous and heavily
context-dependent, {\em e.g.} ``die'' in English could be
a polyhedral random number generator or a transition into death.
Such abstract distortion is not traditionally tackled by 
networking systems (at least not above the relatively
simple problem of bit order and the choice of little or big endian
numeric encodings), though the {\em semantic web} and some content-based
networking schemes\cite{siena:routing,siena:forwarding} have attempted
to address such issues.

\subsection{Fragmentation and Reassembly}
Fragmentation and reassembly arise whenever there is a mismatch
in the desired size of information embodiment and the actual physical
embodiments available. Application-Level Framing (ALF)\cite{alf}
refers to desired application information embodiments as
an {\em Application Data Units}: network devices
have corresponding {\em Transportation Data Units}. Matching
ADUs and TDUs may require significant computation and storage
resources and introduce additional jitter and latency.

\subsection{Multiplexing}
Multiplexing is {\em resource sharing}. Broadly speaking,
resources may be shared in two ways: full access to resources on a
part-time basis or full-time access to reduced resources. 
A realistic multiplexing scheme may employ either or both of these
techniques depending on the nature of the available resources and the
demands of multiplexed clients. When client demands exceed the 
available resources then clients will experience delays (jitter) or
even complete outages. As a further complication, predicting the resource
demands of even {\em one} application may be very difficult, let alone the
cumulative demands of multiple non-cooperating applications, making
it hard to give client applications any realistic guarantees on service.
Multiplexing appears throughout communication systems. TDMA and CSMA
schemes provide part-time access to communication channels, CDMA and OFDM
provide full-time access to reduced communication channels. Circuit-based
systems provide full-time access to a subset of available communication
links, and packet-switched systems provide part-time access to
potentially any communication link in a network. Multiplexing
resources is where social constraints and resources manifest
most obviously, though physical constraints and resources are
clearly of great importance as well.

\subsection{Translation}
Translation is required at semantic boundaries, {\em i.e.}
where there is a change in the {\em meaning}
of symbols as interpreted by some computational entity.
One such example is the Shannon bit boundary: an analog
signal is transformed into a stream of discrete symbols. A rarely-used example
in the canonical network stack is the {\em Presentation} layer, though
it could be argued that the {\em semantic web} functions as a Presentation
layer writ large.

\section{Network System Architecture}
\begin{figure}
\centerline{
\epsfig{file=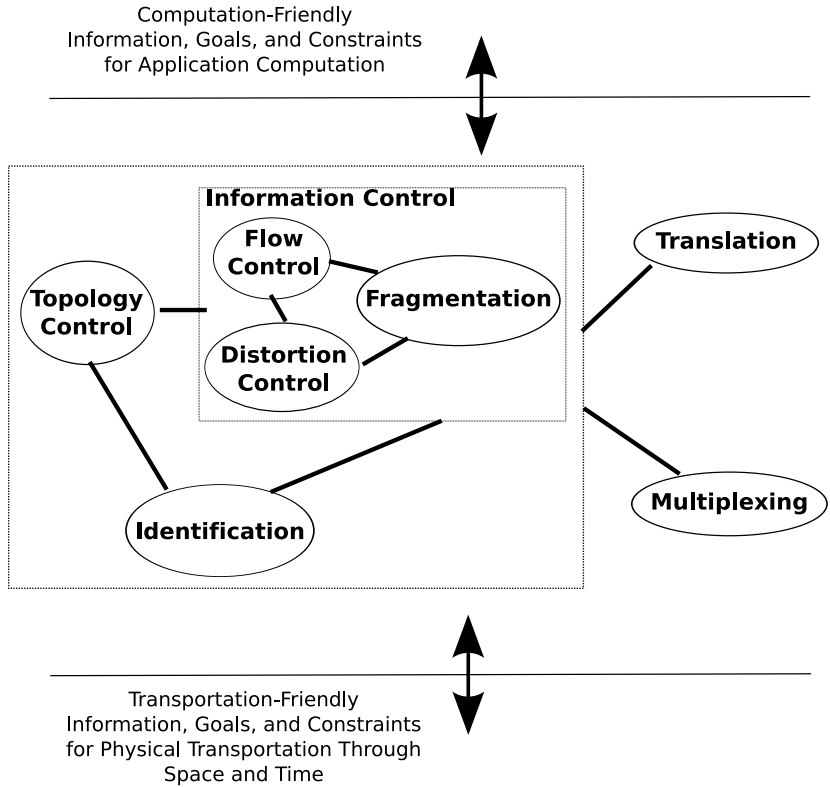,width=0.9\columnwidth}
}
\caption{\footnotesize Interactions between basic network functions. Ovals are used to indicate functions, lines denote potential interactions, and a box indicates a logical grouping of modules. A line connected to a box implies potential interactions with ALL of the individual modules contained within the box.}
\label{figure:network_functions}
\end{figure}

In Figure~\ref{figure:network_functions} we illustrate general interactions
between the seven core network functions we have defined. For clarity we
have grouped Flow Control, Distortion Control, and Fragmentation/Reassembly
together into a single entity called {\em Information Control}.
One of our primary architectural goals is to make and expose the easy
decisions upfront in such a way that specific implementations may
hide the details of the hard decisions from external clients
\cite{parnas-transparency,parnas-modules}, hopefully resulting
in cleaner and more extensible systems. Fortunately,
Figure~\ref{figure:network_functions} offers at least two 
functions that stand out as potential ``easy'' targets:
{\em multiplexing} and {\em translation}.

We can highlight the core nature of multiplexing and translation
to network structure by artificially simplifying network requirements
and functions until only the bare bones of a system remain.
We will make no effort to explicitly identify anything, neither will we
attempt to control or learn about the topology or flow of information.
Our best effort guarantee for the sender will be that information will
travel wherever it can in whatever embodiment is convenient, arriving at
whomever might be listening with an arbitrary and unknown amount of distortion. 
Information may sometimes be fragmented, but never reassembled.
At the receiver we will have a similar guarantee: whatever information
arrives will be passed up to the application for processing as it
arrives. Even this minimalist system would have to perform
at least one task: {\em translation} of information between
embodiments suitable for transport and embodiments suitable for
computation. If we permit multiple applications and/or multiple
communication links our minimalist system would also have to
perform some {\em multiplexing} between those applications
and links. 
We will continue by exploring Translation, Multiplexing, and
the interaction between Topology and Information Control in
greater detail.

\section{Translation and Semantic Boundaries}
\label{section:semanticlayers}
Translation is more obviously layer-friendly than other network
functionality. Information must be translated into a form
that the application can understand before the application
can utilize the information in a meaningful way, and conversely
information from the application must be translated into
a form suitable for transit before the information can be moved.
In this section we will describe how semantic boundaries
may be used to define semantic layers. The bottom semantic
layer hides embodiment-specific details, providing
generic symbol {\em physical transportation} services.
The top semantic layer also hides embodiment-specific details, providing
generic {\em symbolic computation} services where the semantics of the
symbols are determined by individual applications. Strictly speaking, these
two layers are sufficient for system construction. However, it has
proven convenient to consolidate generic symbol transport
services into a third {\em network} layer between computation
and transportation layers. For storage systems one could imagine
a similar stack, with a {\em physical storage} layer replacing the
physical transportation
layer and an intermediate {\em repository} layer replacing
the network layer. A {\em Delay-Tolerant Network} (DTN)
\cite{dtn:rfc} system presents a combined network/repository
layer, perhaps by aggregating separate network/transport and 
repository/storage layers underneath.
\begin{figure}
\centerline{
\epsfig{file=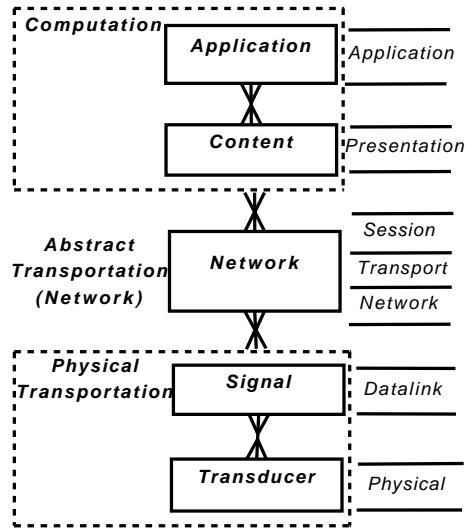,width=0.5\columnwidth}
}
\caption{\footnotesize A picture of our proposed semantic layers. The ``crow's feet'' connectors indicate potential one-to-many relationships for multiplexing. The OSI stack layers are shown on the right side of the diagram for comparison.}
\label{figure:system_meta_architecture}
\end{figure}
Figure~\ref{figure:system_meta_architecture} shows our proposed semantic layer
and sublayer boundaries and how these semantic layers line up with the
layers of the OSI network stack.
We will now describe each of these semantic layers.

\subsection{Physical Transportation Layer}
\label{section:physicallayer}
The {\em Physical Transportation} layer directly manages the motion of symbol
embodiments through space and time and the transfer of symbols to and
from embodiments suitable for transport.
The conversion between transport and computation embodiments
is often (though not always) performed in two main stages.
The {\em Transducer} sublayer transforms the transportation
embodiment into a more computation-friendly embodiment (and vice-versa).
This computation-friendly embodiment is then further processed
by the {\em Signal} layer to transform it into the primary
computation embodiment used by the Network layer. The Signal layer
may itself use multiple computational embodiments as required.
For example, radio communication systems typically use
analog circuits for high-frequency signal processing tasks such
as modulating and demodulating to and from a carrier, but
often opt for more flexible digital systems for processing at baseband.

Practical Transducer and Signal
layers may be very tightly tied together: a particular Signal layer
entity may have to make assumptions about aspects of its attached
Transducer that limit combinations. For example, the sampling rate
used to produce a signal is tightly tied to the physical phenomenon
measured or manipulated by a transducer: attempting to send a 
signal with a gigahertz of bandwidth through an acoustic
transducer and channel may not produce the desired result.
Different systems may have to have vastly different views of
the physical universe. Classical
\footnote{We include electromagnetic waves in the ``classical'' category, even though relativistic and quantum mechanical effects appear.}
waves are sufficient to describe a great many communication systems.
However, quantum mechanical models of the universe are required
for quantum systems\cite{bbnquantum,quantumsearch}, and 
relativity for others\cite{gpsrelativity}. The physical model
of the universe in use and the physical units of captured
signals are important when determining how best to utilize
communication channels. Different systems
that must share a physical channel usually have some shared
models of the channel in order to facilitate joint operation.
This specialized physical knowledge enables the Physical
layer to most effectively negotiate single or multihop
paths through shared physical media. Most existing systems
limit themselves to single hop paths to avoid duplicating
the effort of multihop routing at multiple layers, but
we will not apply a single hop limit at the Physical layer
in our semantic stack. If the Physical layer must utilize
detailed medium-specific knowledge to negotiate multihop paths,
then the task it performs may be {\em different} than the task performed
by multihop routing at higher abstraction layers and does not
constitute duplicated effort. This constraint relaxation should not be taken
as an advocation of rampant and opaque multihop routing at the Physical
layer. In our semantic stack the Physical layer is not in a position
to directly know enough about global traffic demands to make detailed
and autonomous decisions about multihop paths and routing. Such knowledge
is concentrated at the Network layer, and a key job of the Physical layer
is to present the Network layer with transportation options while
attempting to meet the goals and constraints requested by the Network layer.
We will discuss more details of the interactions between the Physical
and Network layers when we discuss the Network layer in
Section~\ref{section:networklayer}.

Concentrating physical concerns down in the Physical layer frees
higher layers from knowledge of specific physical embodiments. However, there
may be some cases where some knowledge of physical embodiments by higher layers
is unavoidable or even desirable.
For example, sensor/actuator applications are inherently physical
in nature, operating ``below the bit'' as analog systems. Even
purely digital systems may also have some coupling to physical
embodiments. A radio communication application might want to
know if a particular segment of a voice transmission was
received on a designated emergency channel in order to alert
the operator. Conversely, if the operator hears a vocal distress
call the operator may want to clear usage of that channel for
emergency traffic, requiring higher layer semantic information
to be pushed back down into the Physical layer. This potential
coupling of higher layer semantics to embodiment-specific
characteristics is a primary place where cross-layer
interactions may enter into our architecture, and we will
discuss this issue in greater detail in
Section~\ref{section:crosslayersemantics}.

\subsection{Computation Layer}
\label{section:computationlayer}
The central {\em networking} tasks for the Computation layer
are to package application symbols into meaningful quanta
({\em i.e.} Application Data Units\cite{alf}),
and to negotiate constraints on distortion, delivery times and
locations for those quanta with the Network layer.
Feedback about available resources from the Network layer may result in
changes to Application Data Unit packaging, {\em e.g.} a streaming
video application might opt to use a lower quality but more compact
encoding scheme if a high bitrate link is not available. We will
discuss more details of interactions between the Computation and
Network layers in the context of the Network layer in
Section~\ref{section:networklayer}.
Multiple applications may opt to consolidate and share information,
and information about application information semantics may be
exploited to shape the flow of information.
Tracking {\em all} possible application symbol semantics is likely
not feasible for a generic networking or storage system. However,
splitting the Computation into two sublayers: the {\em Application}
and {\em Content} layers can facilitate consolidation.

The Application sublayer is equivalent
to the Application layer in common usage today.
The Content sublayer may manage the flow of shared Application
layer information through network and storage systems. The Content layer
could be limited to simple tasks, {\em e.g.} Presentation layer style
conversion of small atomic symbol strings. The Content layer is not
required to have a deep understanding of data semantics.
Traditional network flow control (such as TCP) operates at least partially
within the Content layer: even though the application
data is not being semantically ``understood'' application data is 
summarized and uniquely identified by application type
(port number) and a sequence number. The content-oriented nature
of flow control is emphasized by schemes ({\em e.g.} Structured Streams
\cite{structuredstreams} and SCTP\cite{sctp:rfc}) that permit applications
to specify variable packet delivery requirements than it is
with ``single purpose'' schemes such as TCP and UDP. A Content layer
could also perform complicated wrangling of storage resources,
identification of large aggregate data objects, multiplexing application
usage of underlying communication links, and even encoding content
to better fit available communication resources.
{\em Content-Based Routing} systems\cite{siena:routing,chord} and
content-oriented overlay networks\cite{akamai} illustrate the utility
of having a dedicated Content layer.

\subsection{Network Layer}
\label{section:networklayer}
The {\em Network} layer is the ``center of mass'' in our proposed
architecture. The Network layer must {\em multiplex} applications and
physical network devices, all of which may be competing for limited shared
resources. The Network layer also integrates paths provided by multiple
Physical layer devices and multiple communication endpoints at the ends
of those paths to perform multihop routing. An important aspect
of our semantic stack is that this multiplexing and routing is all
performed within a single layer, instead of three as with the canonical
stack. This unification permits greater flexibility in how these
functions are modularized, rather than forcing layered functionality
for all systems. We would like to emphasize that the layers of
the canonical stack could be implemented within the context of the
semantic stack; such a system could provide an ``in place'' means for 
transitioning between the canonical stack and the
semantic stack. We will not delve into more detail about the
Network layer in this section; many important issues for
the Network layer will be discussed at greater length in
Section~\ref{section:architecturalimplications}.

\section{Architectural Implications}
\label{section:architecturalimplications}
In this section we will highlight some implications of our framework
for system architecture. We will also assess the canonical network
stack and propose techniques for retrofitting existing
layered systems to support ``cross layer'' interactions in a 
systematic way.

\subsection{Cross-Layer Semantic Interactions}
\label{section:crosslayersemantics}
In Section~\ref{section:physicallayer} we briefly discussed the
potential for cross-layer semantic interactions between the Physical
layer and the Computation layer when application-level semantics
relate directly to physical characteristics of an information embodiment.
Sensor/actuator systems are a prime example of such systems, but
are by no means the only one, {\em e.g.} our earlier example
of emergency radio frequency usage. Broadly speaking, we imagine
two ways of approaching this problem: either application-level
semantic information must be pushed into the Physical layer
or physical information must be pushed into the Computation layer.
Both of these approaches couple the Physical and Computation
layers, albeit in different ways. As a general guideline we believe
that when the semantics may be directly tied to the
{\em control plane} (we will further discuss data and control
planes in Section~\ref{section:datacontrolplanes}) the semantics
are suitable for pushing into the Physical and Network layers.
Our ``emergency channel'' example could fall into this category:
an emergency channel is a social construct that may have meaning
for the control plane at all layers of the stack. If the Network
layer also has such a social construct then the ``shortcut'' cross-layer
interaction may be eliminated by propagating information through
the stack in an orderly fashion. A role-based\cite{braden03} approach to
metadata could provide a reasonably flexible mechanism for this task.

On the other hand,
if the information is purely part of the {\em data plane} then it is
suited for pushing up into the Application layer. 
The data plane version of our ``emergency channel'' example
would tag received packets with metadata describing physical properties
and units of the embodiment that contained the packet. In this example
the key pieces of information are the type of embodiment (radio waves),
and the range of frequencies (center and bandwidth) that contained
the information. Further physical information about the radio waves
could include intensity and polarization, and information-theoretic
metrics such as the signal-to-noise ratio could also be of interest.
Such data plane cross layer applications veer strongly into the
realm of generic sensor/actuator systems.
In many ways, sensor/actuator applications reside {\em alongside} the
Physical layer in our semantic stack: we could even imagine
sensor applications sharing a common Transducer layer with the
networking system. These ``parallel'' system structures could become very
confusing, particularly if a sensor/actuator application wishes to use
generic network services. This confusion could be compounded
if the same device used for sensing and actuating were also
used as a network communication device (our emergency channel example
is a very basic instance of a sensor application). We will discuss
the flow of data within the stack (and the potential for
circular flows) in greater detail in Section~\ref{section:datacontrolplanes}.
In general, we will draw a line between sensor systems and network
systems by their {\em intent}. In essence, a networking system
is a very specialized type of sensor/actuator system. If a sensor/actuator
system intends to provide a generic symbolic communication service on top
of some physical embodiment, then architecturally it should reside within 
the Physical layer and not the Application layer. However, it is important
to stress that residing at the Physical layer of a semantic stack
does {\em not} require implementation directly in hardware or
as a kernel module. A time and memory intensive task might be better
placed well outside the kernel in user space, but that placement
does not alter its semantic stack layer. We will discuss such issues
in greater detail in Section~\ref{section:temporalboundaries}.

A further potential semantic cross-layer interaction occurs due
to the fact that many computer systems interact with humans,
and humans (by and large) are resident in the physical world
and are concerned about location, time, and energy. One way
of thinking about this is that user interface systems may
be viewed in terms of semantic layers, connecting the
Application layer with users in yet another Physical layer
({\em e.g.} mice, keyboards, printers, and monitors). Usually
the HCI Physical and Network Physical layers are non-overlapping,
\footnote{Acoustic telephone modems are one rapidly vanishing point at which human and network Physical layers intersect.}
but this shared grounding in the physical world implies that there
may be some utility in placing some physical awareness into the
Network layer. We will discuss this issue in greater detail
in Section~\ref{section:physicalworld}.

\subsection{Routing}
\label{section:routing}
In this section we will highlight some key points about routing
in our semantic stack. We will discuss where and how
multihop paths are determined, how multihop path
information can propagate between layers, and some of the
tradeoffs available when attempting to scalably handle
topology information.

\subsubsection{Multihop Routing and Paths}
Networking systems usually perform the bulk of routing within the Network
layer ({\em e.g.} IP routing) and the Application layer ({\em e.g.}
overlay networks). Routing is also performed at the Physical layer,
but is usually limited to single hop paths (spanning tree bridges
are a common exception). 
An argument made against multihop routing at lower layers is that it leads
to excessive duplication of effort and needless overhead. However,
multihop routing using a particular embodiment within a particular
region of space and time is not necessarily the same thing as 
multihop routing using an abstract graph-theoretic model
\cite{wirelesstopologycontrol}.
Pushing extensive knowledge of the physical layer up into the
network layer is detrimental to overall scalability and abstraction,
and creating a good abstract model of {\em all} physical embodiments
is not a trivial task. It is {\em not} sufficient to simply capture
basic information-theoretic characteristics of an embodiment,
{\em e.g.} in terms of Shannon channel characteristics and interference:
operational characteristics and social norms that apply
to each physical embodiment must be accounted for as well. The potential
complexity of such a task is illustrated by the extensive efforts made
by the Software Defined/Cognitive Radio community\cite{mitola95:sdrarch}
to codify knowledge about the RF domain.

It is important to stress that we are not advocating a model where
lower layers perform isolated, opaque multihop routing and simply present all
reachable endpoints as single-hop neighbors. Such an approach
would hide too much information and control from higher layers.
Instead, we propose that full {\em path} information be made
available to higher layers if requested, and that higher layers
be able to specify requests for desired paths in a way that shapes
how the lower layers establish paths. We will discuss paths
and their representations in greater detail in the next section.

\subsubsection{Generic Path and Routing Metrics}
An important question for layered routing systems is how
to specify route characteristics across layers. This question
is even more important for our semantic stack because of our
emphasis on paths at multiple layers.
Metrics based on fundamental information and physical characteristics
such as symbol capacity, time, space, and energy
seem promising for making abstract routing decisions at the
network/internetwork layers and higher because they relate to
human concerns handled by the Computation layer as well
as embodiment properties handled by the Physical layer.
Such physical metrics are also amenable to graceful adaptation.
For example, instead of forcing a wireless MAC layer to declare a link
``up'' or ``down'' based on some internal quality metric,
the MAC layer could estimate how much time, energy, and space
it would require to move symbols along various paths and let
higher layers determine which paths to use.
{\em Expected Transmission Time} (ETT)\cite{ett} is an example of a
time-based metric which could be extended to include energy and space,
particularly when using
{\em Expected Transmission Count}(ETX)\cite{etx} as a foundation.
Space in this context could be expressed directly in terms of physical
space and/or more abstractly in terms of known endpoints that could receive a
transmission. How the expected values for time, space, and energy
are calculated should be determined by specific Physical layer
entities. For example, if a channel is well-characterized in terms
of its information-theoretic properties then observed SNR
and model predictions of SNR could be used to predict
expected values for path metrics instead of direct feedback-driven
observation of the time, energy, and space required to send packets.

A further critical property of paths is the degree of
{\em entanglement} between different paths.
For example, wireless networks may have
apparently ``independent'' paths that share no endpoints in common but
are mutually interfering. Such entanglement information is critical
for systems that wish to perform multipath routing. The exact
means by which entanglement may be expressed abstractly and succinctly
but with sufficient detail for effective operation is an open issue.
Simple aspects of mutual interference may be expressed through 
degradation of individual path performance when multiple paths are
used. However, actual path {\em independence} may also require
taking physical and social channel characteristics into account. 
For example, two directional free-space optical links might not interfere
with each other at all, but both suffer significant degradation
simultaneously when fog or mist is present. Two optical links
running through the same bundle of fiber would be impervious to fog, but
suffer simultaneous catastrophic failure if the fiber cable were
accidentally cut by construction workers. Such concerns may be difficult
to succinctly and dynamically express. For example, {\em Cognitive Radio}
\cite{mitola-cogradio} systems utilize detailed models and reasoning
in an attempt to capture such physical and social details about
RF systems.

\subsubsection{Scalable Topology Management}
The choices made when designing Information Control and Topology Control 
can have a large effect on system capabilities and resource requirements.
In realistic systems Information Control, Topology Control,
and Identification may be tightly tied together and quite complex.
However, there are some basic tradeoffs that
shape these three primary functions. Systems may opt to place greater
emphasis on either Topology Control or Information Control, and
such a decision can have a significant effect on the demands
made on Identification. 
For example, generic networking systems are designed to support
large quantities and varieties of potentially private information
traversing relatively fewer endpoints. As such it is more practical to
track the endpoints in a network via Topology Control than to
attempt to track all of the information via Information Control.
Conversely, a very dense, specialized sensor network or
special-purpose  ``swarm'' system could have an overwhelming number of
endpoints yet only a few types of content, making it more 
practical to track content instead of worrying about the
locations of particular endpoints.
\begin{figure}
\centerline{
\begin{tabular}{cc}
\psfig{file=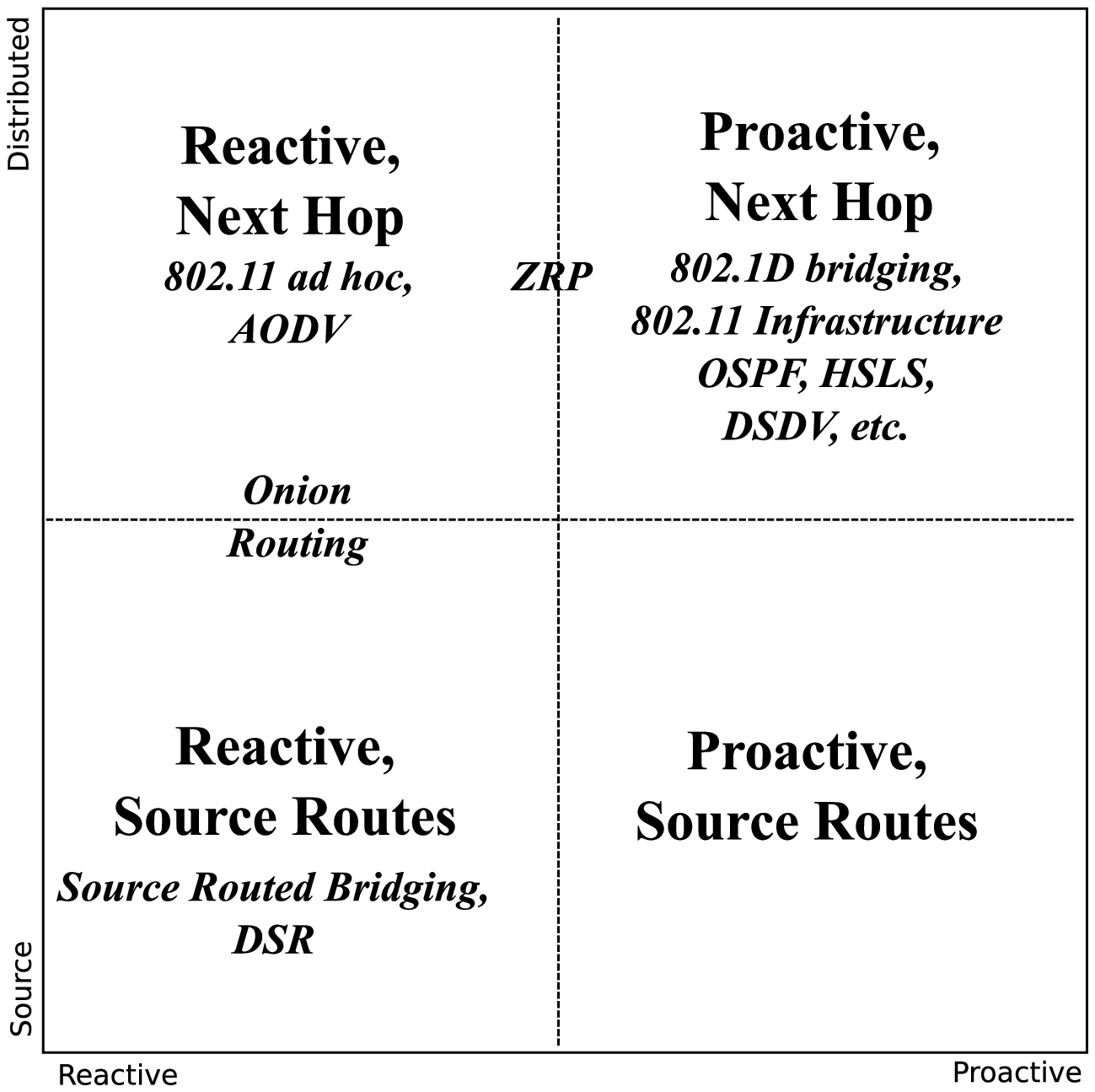,width=0.45\columnwidth}
&
\psfig{file=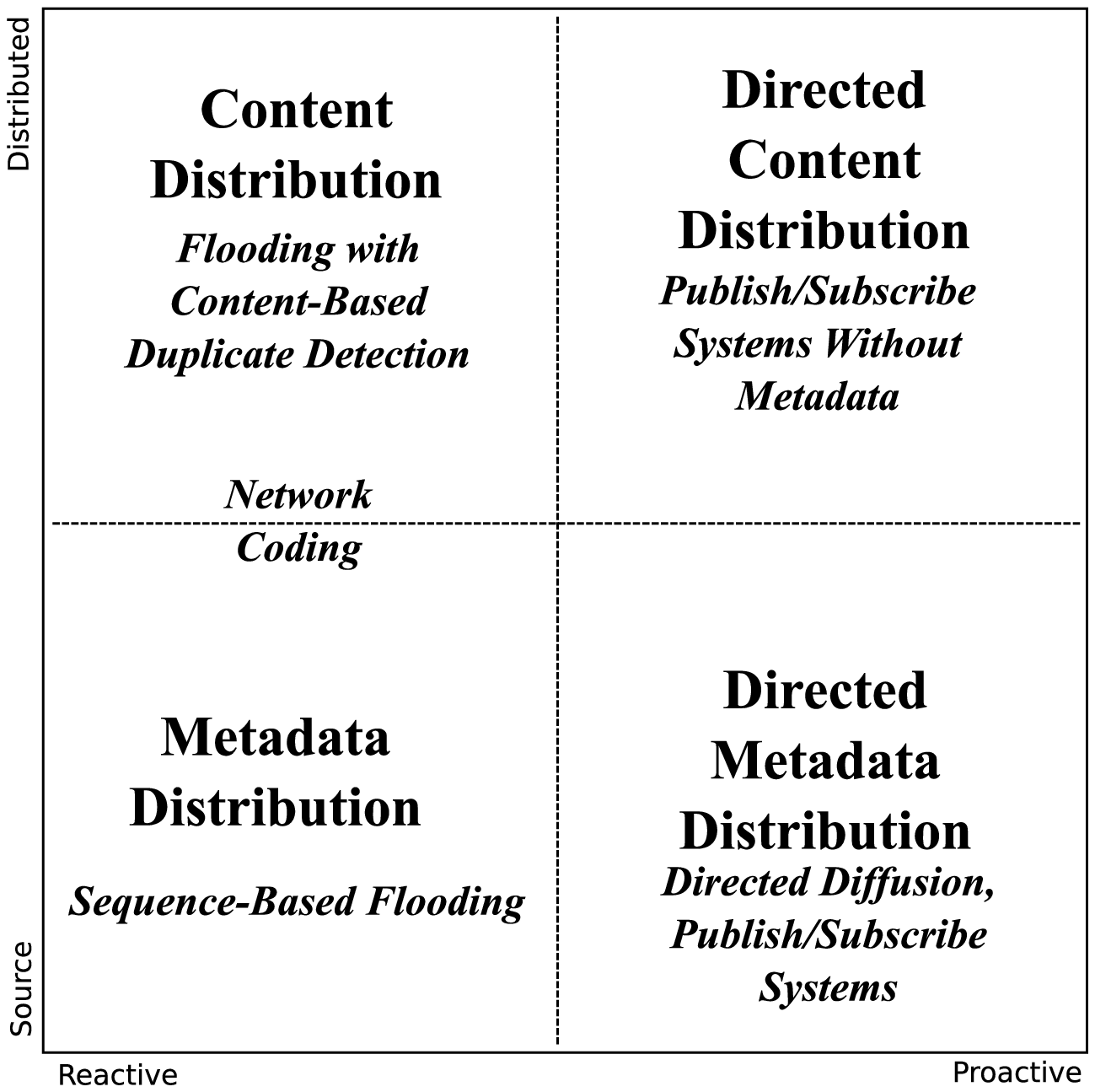,width=0.45\columnwidth}
\\
\begin{minipage}[c]{0.45\columnwidth}
\centerline{
\footnotesize Endpoint-Centric
}
\end{minipage}
&
\begin{minipage}[c]{0.45\columnwidth}
\centerline{
\footnotesize Information-Centric
}
\end{minipage}
\end{tabular}
}
\caption{Routing information may be embedded alongside payload at the {\em source} or computed along the way in a {\em distributed} fashion. Routing decisions may be based within {\em Information Control} and/or {\em Topology Control}, and routing information may be preemptively discovered and maintained or discovered and maintained as required by Flow Control.}
\label{figure:routing_categories}
\end{figure}
Figure~\ref{figure:routing_categories} illustrates some basic tradeoffs
that may be made when constructing network routing systems and makes
a rough assessment of the tradeoffs made by some existing systems.
Systems that require both large numbers of endpoints and
large numbers of pieces of information simultaneously may have
to combine scalable aspects of Information Control and Topology
Control.

\subsection{Flow Control}
In our semantic stack we define a more prominent and distinct role for
Flow Control than is typical. Instead of limiting Flow Control
to specifying {\em when} packets are sent, we also have Flow
Control manage {\em where} packets are sent based on options
provided by Topology Control. This does not entail having Flow
Control calculate shortest-hop paths through the network, but
rather having Flow Control specify topological priorities to 
Topology Control and choosing between paths made available by
Topology Control. Because Flow Control has primary ownership
of packets it would seem to be the entity best suited for multiplexing
usage of limited information pathways provided by Topology Control.
Approaches like Structured Streams\cite{structuredstreams} seem very
promising for letting Flow Control direct Topology Control. Any
number of existing routing protocols are useful for 
Topology Control to provide Flow Control with paths.

\subsection{Data Plane and Control Plane}
\label{section:datacontrolplanes}
Networking systems often utilize separate paths for data and control
information, commonly referred to as the {\em data plane} and
{\em control plane} respectively. The control plane conveys goals,
constraints, actions, and state for topology and information, and the
data plane moves information and associated metadata through
the system for processing. Strictly speaking, the data plane
must always carry some control information. Even if a packet
is handed off with ostensibly {\em no} action requested, there
is an implicit request to {\em store} the packet pending some
future action request, and many systems treat the packet
handoff as an implicit request to {\em transmit} the packet.

The underlying transport and storage resources for control and data
information may be unified, completely separate, or some combination of 
the two. The exact split depends greatly on the resources available
to a system and its requirements. Unified data and control plane
paths are simpler when data and control plane operations are
inherently synchronized, {\em e.g.} a packet handoff contains
an implicit transmit operation and the receipt of a control
packet results in corresponding control plane information.
However, split control/data plane resources could arise
in some situations. A data plane communication
path could be optimized for unidirectional flow of large
data packets alongside a matching control plane optimized for
bidirectional communication of small amounts of
control information. This situation arises for optical networks
that use non-optical IP networks to negotiate the setup and
teardown of high-speed all-optical communication paths.

An interesting property of such optical systems is that
they operate both at the Physical layer by controlling
paths through the optical fabric as well as clients of
generic networking services by using wired IP networks 
as a control plane. This apparent cross-layer interaction
raises questions of how data and control information
flows within a communication stack.
Typical networking stacks maintain straightforward packet flows:
received packets move up the stack until they are consumed
or forwarded back down the stack, and transmitted packets move down the
stack until they are sent or discarded. However, our optical
system example could benefit from relaxation of such flow restrictions.
Instead of creating a cross-layer system (as outlined in 
Section~\ref{section:crosslayersemantics}) to permit operation
both at the Physical and Application layers an alternative
would be to permit the Physical layer to send packets {\em up}
the stack to the adjacent Network layer which would then
deliver the packets as required, sending them back {\em down}
the stack. Care would have to be taken in order
to avoid loops, but with the benefit of averting some potential
cross-layer interactions.

\subsection{Flexible Multiplexing and Temporal Boundaries}
\label{section:temporalboundaries}
\begin{figure}
\centerline{
\epsfig{file=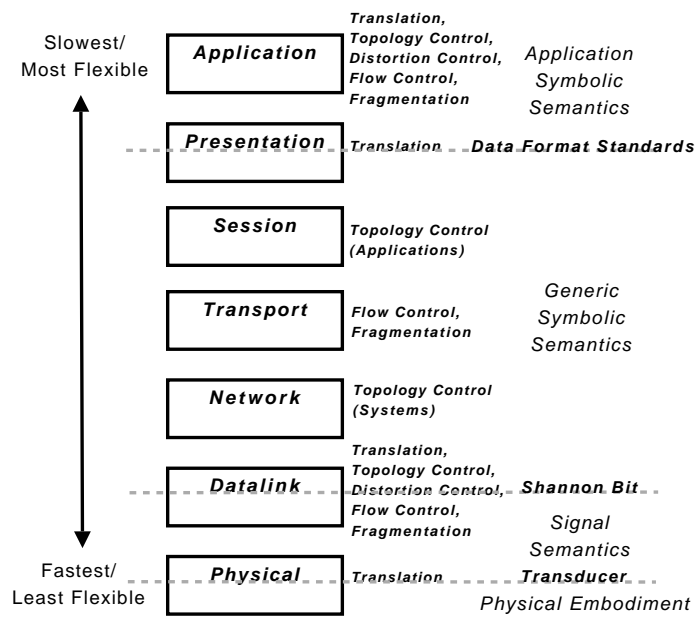,width=0.75\columnwidth}
}
\caption{\footnotesize The OSI stack and its relationship to functional, semantic, and temporal system boundaries. Computational resources are {\em often} (though not always) most flexible and slowest at the top of the stack and least flexible and fastest at the bottom of the stack. {\em Multiplexing} may also be performed by each OSI layer, though the TCP stack does not multiplex to such a large degree.}
\label{figure:osi_stack_commentary}
\end{figure}

Temporal and computational resource constraints have strongly
shaped the canonical network stack. Figure~\ref{figure:osi_stack_commentary}
illustrates the computational flexibility/speed tradeoff implicit in
canonical stack implementations, as well as other functional
and semantic boundaries in the stack.
In our semantic stack we attempt to decouple such temporal and computational
resource constraint tradeoffs from ``height'' in the stack. Each layer
is no longer necessarily monolithically implemented within a single
computational domain, {\em e.g.} as hardware, within the kernel, or
in user space, but instead may be split across different computational
domains. Such splits occur in explicitly in the canonical stack:
the results of routing are memoized within forwarding. The routing/forwarding
split also has benefits for modularity, but modularity on its own does
not require splitting functionality into different layers.
A further example is that of 802.11 MAC layer acknowledgement packets.
Practical implementations may opt to generate MAC layer ACK packets
down alongside the implementation of the PHY layer in order to meet
latency requirements\cite{softmac}.
Dynamically managing computational and temporal constraints, especially
across multiple network hops\cite{clarkrealtime}, is a tricky proposition
in general, and semantic layering on its own does not address
such problems. However, semantic layering does complement systems that
manage such resources by providing larger context and structure.
For example, dynamic flow-based systems such as
{\em Click}\cite{click:tocs00} and {\em Scout}\cite{scoutpaths}
could very easily fit into our semantic layers, as could 
XORP\cite{xorp} and {\em Switchlets}\cite{switchlets}. Such
systems seem well-suited for constructing the ``internal organs''
required for a semantic layer ``exoskeleton'' to come to life.

\subsection{The Physical World}
\label{section:physicalworld}
Some aspects of the physical world are present at all layers of our
semantic stack. We view this as unavoidable because humans are resident
in the physical world and have physical concerns that must be considered
at the Application layer. Combined with the fact that the Physical
layer is concerned with the physical world, and a potential cross-layer
interaction arises, as discussed in Section~\ref{section:crosslayersemantics}.
To help moderate this cross-layer interaction, we propose that the Network
layer have some basic awareness of the physical world. We do not desire
extensive physical knowledge at the Network layer because such knowledge
can erode abstraction. However, we do feel that knowledge of 
{\em time}, {\em space}, and {\em energy} are sufficiently fundamental
that the benefit of adding awareness of them outweighs potential
loss of abstraction. 

Time, space, and energy are by no means new to computation and networking.
In fact, their significance is highlighted by concerted attempts
to create suitable abstractions for them. For example, Lamport
clocks\cite{lamportclocks}, timers in network protocols
such as TCP, geographic routing\cite{grid,lar}, and
more recently in energy-aware systems and programming languages
\cite{eon}. We propose that time, space, and energy should {\em all}
be treated as first-class entities at all layers of the network
stack. It may be difficult (or even impossible) for all systems
to agree on a single shared common notions of space, time, and energy,
but such universal agreement is not required so long as it is possible
to translate between different schemes.
Physical knowledge may sometimes benefit abstraction by placing
absolute bounds on system performance requirements and parameters.
\begin{figure}
\centerline{
\epsfig{file=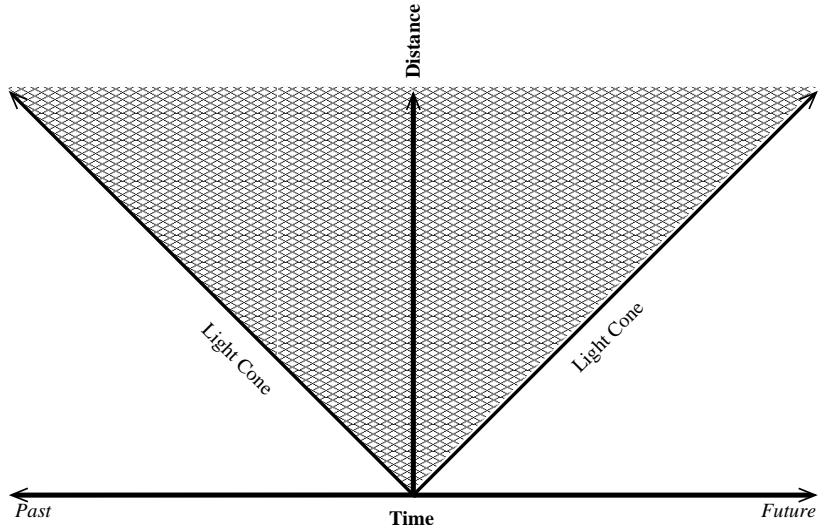,width=0.9\columnwidth}
}
\caption{\footnotesize The light cone of an entity illustrates the maximum range of time and space over which it is physically possible for it to interact based on limits imposed by the speed of light in a vacuum. The regions of space and time filled in by crosshatching are not reachable.}
\label{figure:lightcone}
\end{figure}
For example, the {\em light cone} (shown in Figure~\ref{figure:lightcone})
of an entity shows the largest possible regions of time and space that could
possibly interact with that entity.
At the other end of the scale, the Planck time
and Planck length give theoretical limits on the minimum amounts
of space and time that quantum mechanics may describe, providing
bounds on the absolute resolution required when measuring physical
quantities. Thermodynamics sets basic limits on computational
density and speed\cite{lloyd-1999}. Such bounds may be used to create
``future-proof''  sizes for protocol elements that rely directly
or indirectly on such limits. For example, a software system
that needed to operate over the next 1,000 years could
use a clock value with sufficient bits to track the number
of Planck intervals that occur over 1,000 years if it wanted to be
certain of avoiding clock rollover and/or insufficient precision during
that period of time. Using $365.25$ days per year and $5.39 \times 10^{-44}$
seconds for the Planck time, $179$ bits would be required for
a clock in such a system. Certainly not a small number of bits, but
not unreasonable given that $64$ bit microprocessors are commonly
found in consumer-grade systems. Such concerns are of fundamental
interest to {\em Delay Tolerant Networking} (DTN)\cite{dtn:rfc} 
systems intended for interplanetary and other slow-moving
communication paths.

\subsection{Social Resources and Constraints}
Social concerns arise throughout networking systems and require first-class
design consideration. For example, dynamic spectrum access
systems\cite{dsaoverview} are subject to governmental regulation and
oversight. However, even systems not directly subject to 
government regulation and rules are still replete with 
social norms, albeit usually norms driven by mathematical
models and analysis instead of emerging from autonomous behavior.
An important practical consequence of this pervasive social interaction
is that systems should not simply assume that an ``optimal'' route
in one context is an ``optimal'' route in another context. For example,
a minimum hop path that traverses an untrusted or unfriendly node
may be deemed vastly inferior to a longer path that only traverses
trusted and friendly nodes. Furthermore, different applications
on the same system may adhere to different social norms, potentially
resulting in very different ``optimal'' paths for different applications.
In fact, it is not unreasonable for a single application to have different
social requirements for different pieces of information, resulting in 
even more dynamic behavior. Such operational flexibility clearly has
some computational and architectural complexity cost, but this
tradeoff should be made explicitly. We do not advocate having
a single, unified ``social context'' for {\em all} layers of the stack
unless such a context truly exists. One great benefit
of layering, abstraction, and delegation is that local details may be 
kept local.

%
%
%
%
%
%
%

\section{Summary and Future Work}
In this work we have presented a layered network architecture
based on {\em semantics}. We have justified this structure
using reasoning based on physics, information theory, software engineering
principles, and aspects of human/computer interaction. Naturally, 
there are many details of such a system to explore. We plan to move forward
by designing and constructing practical systems based on our architecture,
utilizing well-established and compatible platforms,
{\em e.g.} the Click Modular Router\cite{click:tocs00} and
XORP\cite{xorp}, to fill out the framework laid out by our
semantic stack. As an initial primary point of exploration we are
looking at paths and path representations. Paths and path properties
are a central part of interactions between layers in our architecture,
and as such we feel they warrant detailed exploration.

\bibliographystyle{unsrt}
\bibliography{semanticlayering}

\begin{thebibliography}{10}

\bibitem{osi:zimmerman}
Hubert Zimmerman.
\newblock {OSI} reference model - the {ISO} model of architecture for open
  systems interconnection.
\newblock {\em IEEE Transactions on Communications}, 28(4):425--432, April
  1980.

\bibitem{feynman}
Richard~Phillips Feynman.
\newblock {\em Feynman Lectures on Computation}.
\newblock Perseus Books, Cambridge, MA, USA, 2000.

\bibitem{zuse}
Konrad Zuse.
\newblock Rechnender raum.
\newblock In {\em Physik und Informatik - Informatik und Physik,
  Arbeitsgespr\"{a}ch}, pages 16--23, London, UK, 1992. Springer-Verlag.

\bibitem{fredkin}
E.~Fredkin.
\newblock Digital mechanics: an informational process based on reversible
  universal cellular automata.
\newblock {\em Phys. D}, 45(1-3):254--270, 1990.

\bibitem{denotationalsemantics}
Joseph~E. Stoy.
\newblock {\em Denotational Semantics: The Scott-Strachey Approach to
  Programming Language Theory}.
\newblock MIT Press, Cambridge, MA, USA, 1981.

\bibitem{shannon-bit}
C.~E. Shannon.
\newblock A mathematical theory of communication.
\newblock {\em Bell System Technical Journal}, 27:379--423,623--656,
  July/October 1948.

\bibitem{kleinrock1}
L.~Kleinrock and F.~Tobagi.
\newblock Packet switching in radio channels: Part i--carrier sense
  multiple-access modes and their throughput-delay characteristics.
\newblock {\em Communications, IEEE Transactions on [legacy, pre - 1988]},
  23(12):1400--1416, Dec 1975.

\bibitem{dsaoverview}
Qing Zhao and Brian~M. Sadler.
\newblock Dynamic spectrum access: Signal processing, networking, and
  regulatory policy, 2006.

\bibitem{netmotion}
P.~Basu and J.~Redi.
\newblock Movement control algorithms for realization of fault-tolerant ad hoc
  robot networks.
\newblock {\em Network, IEEE}, 18(4):36--44, July-Aug. 2004.

\bibitem{molecular}
M.~Moore., A.~Enomoto, T.~Nakano, R.~Egashira, T.~Suda, A.~Kayasuga, H.~Kojima,
  H.~Sakakibara, and K.~Oiwa.
\newblock A design of a molecular communication system for nanomachines using
  molecular motors.
\newblock In {\em Proceedings of the Fourth Annual IEEE Conference on Pervasive
  Computing and Communications and Workshops}, March 2006.

\bibitem{kleinrock2}
F.~Tobagi and L.~Kleinrock.
\newblock Packet switching in radio channels: Part ii--the hidden terminal
  problem in carrier sense multiple-access and the busy-tone solution.
\newblock {\em Communications, IEEE Transactions on [legacy, pre - 1988]},
  23(12):1417--1433, Dec 1975.

\bibitem{80211spec}
{W}ireless {LAN} {M}edium {A}ccess {C}ontrol ({MAC}) and {P}hysical {L}ayer
  ({PHY}) {Spec}, {IEEE} 802.11 {Standard}.
\newblock Technical report, Institute of Electrical and Electronics Engineers,
  Inc.

\bibitem{alf}
D.~D. Clark and D.~L. Tennenhouse.
\newblock Architectural considerations for a new generation of protocols.
\newblock In {\em SIGCOMM '90: Proceedings of the ACM symposium on
  Communications architectures \& protocols}, pages 200--208, New York, NY,
  USA, 1990. ACM Press.

\bibitem{arch:rfc}
R.~Bush and D.~Meyer.
\newblock {Some Internet Architectural Guidelines and Philosophy}.
\newblock {IETF RFC} 3439, {IETF}, December 2002.

\bibitem{names-hauzeur}
Bernard~M. Hauzeur.
\newblock A model for naming, addressing and routing.
\newblock {\em ACM Trans. Inf. Syst.}, 4(4):293--311, 1986.

\bibitem{mobileip:rfc}
Charlie Perkins.
\newblock {IP Mobility Support}.
\newblock {IETF RFC} 2002, {IETF}, October 1996.

\bibitem{sctp-multipath}
Janardhan~R. Iyengar, Paul~D. Amer, and Randall Stewart.
\newblock Concurrent multipath transfer using sctp multihoming over independent
  end-to-end paths.
\newblock {\em IEEE/ACM Trans. Netw.}, 14(5):951--964, 2006.

\bibitem{siena:routing}
Antonio Carzaniga, Matthew~J. Rutherford, and Alexander~L. Wolf.
\newblock A routing scheme for content-based networking.
\newblock In {\em Proceedings of IEEE INFOCOM 2004}, Hong Kong, China, March
  2004.

\bibitem{siena:forwarding}
Antonio Carzaniga and Alexander~L. Wolf.
\newblock Forwarding in a content-based network.
\newblock In {\em Proceedings of ACM SIGCOMM 2003}, pages 163--174, Karlsruhe,
  Germany, August 2003.

\bibitem{parnas-transparency}
D.~L. Parnas and D.~P. Siewiorek.
\newblock Use of the concept of transparency in the design of hierarchically
  structured systems.
\newblock {\em Commun. ACM}, 18(7):401--408, 1975.

\bibitem{parnas-modules}
D.~L. Parnas.
\newblock On the criteria to be used in decomposing systems into modules.
\newblock pages 139--150, 1979.

\bibitem{dtn:rfc}
V.~Cerf, S.~Burleigh, A.~Hooke, L.~Torgerson, R.~Durst, K.~Scott, K.~Fall, and
  H.~Weiss.
\newblock {Delay-Tolerant Networking Architecture}.
\newblock {IETF RFC} 4838, {IETF}, April 2007.

\bibitem{bbnquantum}
Chip Elliott, David Pearson, and Gregory Troxel.
\newblock Quantum cryptography in practice.
\newblock In {\em SIGCOMM '03: Proceedings of the 2003 conference on
  Applications, technologies, architectures, and protocols for computer
  communications}, pages 227--238, New York, NY, USA, 2003. ACM Press.

\bibitem{quantumsearch}
Lov~K. Grover.
\newblock A fast quantum mechanical algorithm for database search.
\newblock In {\em STOC '96: Proceedings of the twenty-eighth annual ACM
  symposium on Theory of computing}, pages 212--219, New York, NY, USA, 1996.
  ACM.

\bibitem{gpsrelativity}
Neil Ashby.
\newblock Relativity in the global positioning system.
\newblock {\em Living Reviews in Relativity}, 6(1), 2003.

\bibitem{structuredstreams}
Bryan Ford.
\newblock Structured streams: a new transport abstraction.
\newblock {\em SIGCOMM Comput. Commun. Rev.}, 37(4):361--372, 2007.

\bibitem{sctp:rfc}
R.~Stewart, Q.~Xie, K.~Morneault, C.~Sharp, H.~Schwarzbauer, T.~Taylor,
  I.~Rytina, M.~Kalla, L.~Zhang, and V.~Paxson.
\newblock {Stream Control Transmission Protocol}.
\newblock {IETF RFC} 2960, {IETF}, October 2000.

\bibitem{chord}
Ion Stoica, Robert Morris, David Karger, M.~Frans Kaashoek, and Hari
  Balakrishnan.
\newblock Chord: A scalable peer-to-peer lookup service for internet
  applications.
\newblock In {\em SIGCOMM '01: Proceedings of the 2001 conference on
  Applications, technologies, architectures, and protocols for computer
  communications}, pages 149--160, New York, NY, USA, 2001. ACM.

\bibitem{akamai}
{Akamai, Inc.}
\newblock {Akamai}.
\newblock \verb|http://www.akamai.com|.

\bibitem{braden03}
Robert Braden, Ted Faber, and Mark Handley.
\newblock From protocol stack to protocol heap: role-based architecture.
\newblock {\em SIGCOMM Comput. Commun. Rev.}, 33(1):17--22, 2003.

\bibitem{wirelesstopologycontrol}
Yan Gao, J.C. Hou, and Hoang Nguyen.
\newblock Topology control for maintaining network connectivity and maximizing
  network capacity under the physical model.
\newblock {\em INFOCOM 2008. The 27th Conference on Computer Communications.
  IEEE}, pages 1013--1021, April 2008.

\bibitem{mitola95:sdrarch}
Joe Mitola.
\newblock The software radio architecture.
\newblock {\em IEEE Communications Magazine}, 33(5):26--38, May 1995.

\bibitem{ett}
Richard Draves, Jitendra Padhye, and Brian Zill.
\newblock Routing in multi-radio, multi-hop wireless mesh networks.
\newblock In {\em MobiCom '04: Proceedings of the 10th annual international
  conference on Mobile computing and networking}, pages 114--128, New York, NY,
  USA, 2004. ACM.

\bibitem{etx}
Douglas S. J.~De Couto, Daniel Aguayo, John Bicket, and Robert Morris.
\newblock A high-throughput path metric for multi-hop wireless routing.
\newblock {\em Wirel. Netw.}, 11(4):419--434, 2005.

\bibitem{mitola-cogradio}
III Mitola, J. and Jr. Maguire, G.Q.
\newblock Cognitive radio: making software radios more personal.
\newblock {\em Personal Communications, IEEE}, 6(4):13--18, Aug 1999.

\bibitem{softmac}
Michael Neufeld, Jeff Fifield, Christian Doerr, Anmol Sheth, and Dirk Grunwald.
\newblock Softmac - flexible wireless research platform.
\newblock In {\em Fourth Workshop on Hot Topics in Networks (HotNets-IV)},
  November 2005.

\bibitem{clarkrealtime}
David~D. Clark, Scott Shenker, and Lixia Zhang.
\newblock Supporting real-time applications in an integrated services packet
  network: architecture and mechanism.
\newblock {\em SIGCOMM Comput. Commun. Rev.}, 22(4):14--26, 1992.

\bibitem{click:tocs00}
Eddie Kohler, Robert Morris, Benjie Chen, John Jannotti, and M.~Frans Kaashoek.
\newblock The click modular router.
\newblock {\em {ACM} Transactions on Computer Systems}, 18(3):263--297, August
  2000.

\bibitem{scoutpaths}
David Mosberger and Larry~L. Peterson.
\newblock Making paths explicit in the scout operating system.
\newblock pages 153--167, 1996.

\bibitem{xorp}
Mark Handley, Orion Hodson, and Eddie Kohler.
\newblock Xorp: an open platform for network research.
\newblock {\em SIGCOMM Comput. Commun. Rev.}, 33(1):53--57, 2003.

\bibitem{switchlets}
J.~E. Van~Der Merwe and I.~M. Leslie.
\newblock Switchlets and dynamic virtual atm networks.
\newblock In {\em Proc Integrated Network Management V}, pages 355--368.
  Chapman and Hall, 1997.

\bibitem{lamportclocks}
Leslie Lamport.
\newblock Time, clocks, and the ordering of events in a distributed system.
\newblock {\em Commun. ACM}, 21(7):558--565, 1978.

\bibitem{grid}
J.~Li, J.~Jannotti, D.~{De Couto}, D.~Karger, and R.~Morris.
\newblock A scalable location service for geographic ad-hoc routing.
\newblock In {\em Proceedings of the 6th {ACM} International Conference on
  Mobile Computing and Networking ({MobiCom} '00)}, pages 120--130, August
  2000.

\bibitem{lar}
Young-Bae Ko and Nitin~H. Vaidya.
\newblock Location-aided routing (lar) in mobile ad hoc networks.
\newblock {\em Wireless Networks}, 6(4):307--321, 2000.

\bibitem{eon}
Jacob Sorber, Alexander Kostadinov, Matthew Garber, Matthew Brennan, Mark~D.
  Corner, and Emery~D. Berger.
\newblock Eon: a language and runtime system for perpetual systems.
\newblock In {\em SenSys '07: Proceedings of the 5th international conference
  on Embedded networked sensor systems}, pages 161--174, New York, NY, USA,
  2007. ACM.

\bibitem{lloyd-1999}
Seth Lloyd.
\newblock Ultimate physical limits to computation, 1999.

\end{thebibliography}

\end{document}